\documentclass[aps,twocolumn,showpacs,preprintnumbers,amsmath,amssymb,superscriptaddress,floatfix,nofootinbib]{revtex4}

\usepackage{graphicx}
\usepackage{hyperref}
\usepackage{amsmath}
\allowdisplaybreaks[1]
\usepackage{amsfonts}
\usepackage{amssymb}
\usepackage{multirow}
\usepackage{makecell}
\usepackage[dvipsnames]{xcolor}
\usepackage{booktabs}
\usepackage{tabularx}
\usepackage{subfigure}
\usepackage{makecell}
\usepackage{ulem}

\newcommand{\Slash}[1]{\ooalign{\hfil/\hfil\crcr$#1$}}

\begin{document}


\title{Resonances in heavy meson - heavy baryon coupled-channel interactions}

\author{Zheng-Li Wang} \email{wangzhengli@itp.ac.cn}
\affiliation{CAS Key Laboratory of Theoretical Physics, Institute of Theoretical Physics, \\
Chinese Academy of Sciences, Beijing 100190,China}
\affiliation{School of Physics, University of Chinese Academy of Sciences (UCAS), Beijing 100049, China}

\author{Chao-Wei Shen} \email{c.shen@fz-juelich.de}
\affiliation{Institute for Advanced Simulation, Institut f\"ur Kernphysik and
J\"ulich Center for Hadron Physics, \\ Forschungszentrum J\"ulich, D-52425 J\"ulich, Germany}

\author{Deborah R\"onchen} \email{d.roenchen@fz-juelich.de}
\affiliation{Institute for Advanced Simulation, Institut f\"ur Kernphysik and
J\"ulich Center for Hadron Physics, \\ Forschungszentrum J\"ulich, D-52425 J\"ulich, Germany}

\author{Ulf-G. Mei\ss{}ner} \email{meissner@hiskp.uni-bonn.de}
\affiliation{Helmholtz-Institut f\"ur Strahlen- und Kernphysik and Bethe Center for
Theoretical Physics,\\ Universit\"at Bonn, D-53115 Bonn, Germany}
\affiliation{Institute for Advanced Simulation, Institut f\"ur Kernphysik and J\"ulich
Center for Hadron Physics, \\ Forschungszentrum J\"ulich, D-52425 J\"ulich, Germany}
\affiliation{Tbilisi State University, 0186 Tbilisi, Georgia}

\author{Bing-Song Zou} \email{zoubs@itp.ac.cn}
\affiliation{CAS Key Laboratory of Theoretical Physics, Institute of Theoretical Physics, \\
Chinese Academy of Sciences, Beijing 100190,China}
\affiliation{School of Physics, University of Chinese Academy of Sciences (UCAS), Beijing 100049, China}
\affiliation{School of Physics and Electronics, Central South University, Changsha 410083, China}

\date{\today}

\begin{abstract}
The interactions of $\bar{D}^{(*)} \Lambda_c - \bar{D}^{(*)}\Sigma_c^{(*)}$ are studied within
the framework of a dynamical coupled-channel approach. A series of bound states and resonances with different spin and parity are dynamically generated
in the hidden charm sector. Four $S$-wave bound states are found in the
mass range of 4.3 to 4.5~GeV, close to the pentaquark states observed by LHCb. Two of the states have a spin parity of $J^P= 1/2^-$ and the other two
have $J^P=3/2^-$. In addition, several resonances with different spin and parity in higher partial waves are predicted.
\end{abstract}

\maketitle

\section{Introduction}

In 2015, two pentaquark-like resonances were observed by the LHCb Collaboration in the $J/\psi p$
invariant mass spectrum of the $\Lambda_b^0 \to J/\psi p K^-$ decay~\cite{LHCb:2015yax}.
The results of these $P_c$ states were then updated in 2019 with three clear narrow structures
reported in Ref.~\cite{LHCb:2019kea}.
Recently, another new pentaquark state $P_c(4337)$ has been announced with a significance
between 3.1 to $3.7\sigma$ by the LHCb Collaboration~\cite{LHCb:2021chn}.
The existence of these pentaquark-like states with hidden charm had been predicted before
the experimental findings in Refs.~\cite{Wu:2010jy,Wang:2011rga,Yang:2011wz,Yuan:2012wz,Wu:2012md,Xiao:2013yca,Uchino:2015uha}, and they attracted much attention after the observation.
Among the theoretical works, a popular explanation is the anticharmed meson-charmed baryon
molecular picture, where these exotic states are $\bar{D}^{(*)} \Sigma_c^{(*)}$
bound states~\cite{Oset:2012ap,
Garcia-Recio:2013gaa,Chen:2015loa,Chen:2015moa,Roca:2015dva,
He:2015cea,Meissner:2015mza,Huang:2015uda,Wang:2015qlf,Yang:2015bmv,Chen:2016heh,
Roca:2016tdh,Lu:2016nnt,Shimizu:2016rrd,Shen:2016tzq,Ortega:2016syt,Yamaguchi:2016ote,
He:2016pfa,Oset:2016nvf,Lin:2017mtz,Yamaguchi:2017zmn,Shen:2017ayv,Lin:2018kcc,
Liu:2019tjn,Xiao:2019aya,Du:2019pij,Yalikun:2021bfm}.

In Ref.~\cite{Shen:2017ayv}, the authors applied the J\"ulich-Bonn dynamical coupled-channel
(J\"uBo DCC) framework~\cite{Ronchen:2012eg} to the hidden-charm sector to investigate the resonance spectrum in
the $\bar{D}\Lambda_c - \bar{D}\Sigma_c$ system.
The J\"uBo approach was originally developed to extract the light baryon resonance spectrum from $\pi N$ scattering data and has been successfully applied in the simultaneous analysis of pion- and photon-induced reactions~\cite{Ronchen:2014cna,Ronchen:2015vfa,Ronchen:2018ury}. Recently, it was extended to include also electroproduction reactions~\cite{Mai:2021vsw,Mai:2021aui}.   The scattering amplitude is obtained as the solution of a Lippmann-Schwinger
 equation respecting theoretical constraints on the $S$-matrix like unitarity and analyticity.
%
%
In the exploratory study of Ref.~\cite{Shen:2017ayv}, only two channels $\bar{D}\Lambda_c$ and $\bar{D}\Sigma_c$ with
vector meson exchange in the $t$-channel were included and one pole in each partial wave up to
$J^P=5/2^+$ and $5/2^-$ was dynamically generated. For $J^P=1/2^-$, there exist a very narrow pole
around $4296$~MeV, which can be considered as a $\bar{D}\Sigma_c$ bound state.

Here, we extend the work of Ref.~\cite{Shen:2017ayv} to a more complete coupled-channel
calculation, including $\bar{D}\Lambda_c$, $\bar{D}\Sigma_c$, $\bar{D}^* \Lambda_c$, $\bar{D}^*
\Sigma_c$, $\bar{D} \Sigma_c^*$ channels, where $t$-channel pseudoscalar and vector meson exchange
and $u$-channel doubly charmed baryon exchange are taken into account. No ``genuine" $s$-channel poles are included.
%
%
Possible dynamically generated poles in different partial
waves in the  energy range around 4~GeV are studied. Together with their couplings to the different channels,
we expect to obtain a more comprehensive picture of the hidden-charm resonance spectrum.

This work is organized as follows.
In Sec.~\ref{Sec:theory}, we present the theoretical framework of our calculation.
In Sec.~\ref{Sec:result}, the numerical results for the $\bar{D}\Lambda_c$, $\bar{D}\Sigma_c$,
$\bar{D}^* \Lambda_c$, $\bar{D}^* \Sigma_c$, $\bar{D} \Sigma_c^*$ interactions and relevant
discussions are presented.
We end with a brief summary in Sec.~\ref{Sec:summary}. Many technicalities are relegated to the
Appendix.
%

\section{Formalism} \label{Sec:theory}

In the J\"uBo model, the scattering equation describing the coupled-channel interactions for a given partial wave
takes the form
\begin{align} \label{Eq:LSEq}
    T_{\mu\nu}(p^{\prime\prime}, p^\prime, z) =& V_{\mu\nu}(p^{\prime\prime}, p^\prime, z) +
        \sum_\kappa \int_0^\infty {\rm d}p p^2 V_{\mu\kappa}(p^{\prime\prime}, p, z) \nonumber \\
        & \times G_\kappa(p, z) T_{\kappa\nu}(p, p^\prime, z),
\end{align}
where $z$ is the scattering energy in the center-of-mass system, $p^{\prime\prime}=|\vec{p^{\prime\prime}}|$,
$p^{\prime}=|\vec{p^{\prime}}|$ and $p=|\vec{p}|$ represent the out-going, in-coming, and intermediate
three-momenta, respectively, which may be on- or off-shell.
%
%
The subscripts $\mu$, $\nu$, $\kappa$ are channel indices, and the integral term contains a sum over all intermediate possible quantum numbers and channels involved in the model.
The propagator $G_\kappa(p, z)$ is given by
\begin{align}
    G_\kappa(p, z) = \frac1{z-E_a(p)-E_b(p)+i\varepsilon},
\end{align}
where $E_a(p)=\sqrt{m_a^2+p^2}$ and $E_b(p)=\sqrt{m_b^2+p^2}$ are the on-shell energies of the
intermediate particles $a$ and $b$ in the channel $\kappa$.
In Eq.~\eqref{Eq:LSEq} left-hand cuts and the correct structure of complex branch points are implemented, which allows for a well-defined determination of resonances as poles in the complex energy plane on the second Riemann sheet, see Ref.~\cite{Doring:2009yv} for a detailed discussion.

%
%
It should be pointed out that there are different possibilities for the $\bar{D}^* \Lambda_c$,
$\bar{D}^* \Sigma_c$ and $\bar{D} \Sigma_c^*$ channels to couple to a given spin-parity $J^P$.
Different coupling possibilities for these channels are denoted as different channels $\mu$.
In Table~\ref{Tab:QuNum}, we present the complete coupling scheme utilized in this work,
where $S$ is the total spin and $L$ is the orbital angular momentum.
The notation of the quantum numbers is the common spectroscopic notation, $L_{2I,2J}$, with $I$ denoting the isospin.
%
%
%

\begin{table*}[htbp]
    \centering
    \renewcommand\arraystretch{1.5}
    \caption{Angular momentum structure of the coupled channels in isospin $I=\frac12$ up to $J=\frac92$. \label{Tab:QuNum}}
    \begin{tabular}{p{0.5cm}<{\centering}l|*{2}{p{0.8cm}<{\centering}}|*{2}{p{0.8cm}<{\centering}}|*{2}{p{0.8cm}<{\centering}}|*{2}{p{0.8cm}<{\centering}}|*{2}{p{0.8cm}<{\centering}}}
        \hline
        \hline
        $\mu$ & \qquad \qquad \qquad \qquad \qquad \quad $J^P=$ & $\frac12^-$ & $\frac12^+$ & $\frac32^+$ & $\frac32^-$ & $\frac52^-$ & $\frac52^+$ & $\frac72^+$ & $\frac72^-$ & $\frac{9}{2}^-$ & $\frac{9}{2}^+$ \\
        \hline
        1 & $\bar{D}\Lambda_c$ & $S_{11}$ & $P_{11}$ & $P_{13}$ & $D_{13}$ & $D_{15}$ & $F_{15}$ & $F_{17}$ & $G_{17}$ & $G_{19}$ & $H_{19}$\\
        2 & $\bar{D}\Sigma_c$ & $S_{11}$ & $P_{11}$ & $P_{13}$ & $D_{13}$ & $D_{15}$ & $F_{15}$ & $F_{17}$ & $G_{17}$ & $G_{19}$ & $H_{19}$\\
        3 & $\bar{D}^* \Lambda_c(S=1/2)$ & $S_{11}$ & $P_{11}$ & $P_{13}$ & $D_{13}$ & $D_{15}$ & $F_{15}$ & $F_{17}$ & $G_{17}$ & $G_{19}$ & $H_{19}$\\
        4 & $\bar{D}^* \Lambda_c(S=3/2, |J-L|=1/2)$ & - & $P_{11}$ & $P_{13}$ & $D_{13}$ & $D_{15}$ & $F_{15}$ & $F_{17}$ & $G_{17}$ & $G_{19}$ & $H_{19}$\\
        5 & $\bar{D}^* \Lambda_c(S=3/2, |J-L|=3/2)$ & $D_{11}$ & - & $F_{13}$ & $S_{13}$ & $G_{15}$ & $P_{15}$ & $H_{17}$ & $D_{17}$ & $I_{19}$ & $F_{19}$\\
        6 & $\bar{D}^* \Sigma_c(S=1/2)$ & $S_{11}$ & $P_{11}$ & $P_{13}$ & $D_{13}$ & $D_{15}$ & $F_{15}$ & $F_{17}$ & $G_{17}$ & $G_{19}$ & $H_{19}$\\
        7 & $\bar{D}^* \Sigma_c(S=3/2, |J-L|=1/2)$ & - & $P_{11}$ & $P_{13}$ & $D_{13}$ & $D_{15}$ & $F_{15}$ & $F_{17}$ & $G_{17}$ & $G_{19}$ & $H_{19}$\\
        8 & $\bar{D}^* \Sigma_c(S=3/2, |J-L|=3/2)$ & $D_{11}$ & - & $F_{13}$ & $S_{13}$ & $G_{15}$ & $P_{15}$ & $H_{17}$ & $D_{17}$ & $I_{19}$ & $F_{19}$\\
        9 & $\bar{D} \Sigma_c^*(|J-L|=1/2)$ & - & $P_{11}$ & $P_{13}$ & $D_{13}$ & $D_{15}$ & $F_{15}$ & $F_{17}$ & $G_{17}$ & $G_{19}$ & $H_{19}$\\
        10 & $\bar{D} \Sigma_c^*(|J-L|=3/2)$ & $D_{11}$ & - & $F_{13}$ & $S_{13}$ & $G_{15}$ & $P_{15}$ & $H_{17}$ & $D_{17}$ & $I_{19}$ & $F_{19}$\\
        \hline
        \hline
    \end{tabular}
\end{table*}

The potential $V$ iterated in Eq.\eqref{Eq:LSEq} is constructed from the effective interactions
based on the Lagrangians of Wess and Zumino~\cite{Wess:1967jq,Meissner:1987ge}, using
time-ordered perturbation theory (TOPT).
The involved effective Lagrangians are
\begin{align}
	{\cal L}_{PPV} =& i \sqrt2 g_{PPV} (P \partial^\mu P - \partial^\mu P P) V_\mu,	\nonumber \\
	{\cal L}_{VVP} =& \frac{g_{VVP}}{m_V} \epsilon_{\mu\nu\alpha\beta} \partial^\mu V^\nu \partial^\alpha V^\beta P, \nonumber \\
	{\cal L}_{VVV} =& i g_{VVV} \langle V^\mu [V^\nu, \partial_\mu V_\nu] \rangle,	\nonumber \\
	{\cal L}_{BBP} =& \frac{g_{BBP}}{m_P} \bar{B} \gamma^\mu \gamma^5 \partial_\mu P B,	\nonumber \\
	{\cal L}_{BBV} =& - g_{BBV} \bar{B} (\gamma^\mu - \frac{\kappa}{2m_B} \sigma^{\mu\nu} \partial_\nu) V_\mu B,	\nonumber \\
	{\cal L}_{BDP} =& - \frac{g_{BDP}}{m_P} (\bar{B} \partial^\mu P D_\mu + \bar{D}_\mu \partial^\mu P B),	\nonumber \\
	{\cal L}_{BDV} =& i \frac{g_{BDV}}{m_V} \big[ \bar{B} \gamma_\mu \gamma^5 D_\nu (\partial^\mu V^\nu - \partial^\nu V^\mu) \nonumber \\
        & + \bar{D}_\mu \gamma_\nu \gamma^5 B (\partial^\mu V^\nu - \partial^\nu V^\mu) \big],	\nonumber \\
	{\cal L}_{DDV} =& g_{DDV} \bar{D}^\tau (\gamma^\mu - \frac{\kappa}{2m_D} \sigma^{\mu\nu} \partial_\nu) V_\mu D_\tau,
\end{align}
where $P$, $V$, $B$ and $D$ denote the pseudoscalar, vector meson, and baryon octet, in order,
and the decuplet baryons.
We have extended the original formulation~\cite{Ronchen:2012eg} for three flavors to four, but of course this
flavor SU(4) symmetry is strongly broken by the meson and baryon masses.
We apply SU(3) and SU(4) flavor symmetry~\cite{deSwart:1963pdg,Okubo:1975sc} to relate the specific coupling constants.
The expressions of the amplitudes, detailed relations of the couplings and other theoretical
formulae are given in Appendix~\ref{App:formalism}.
%

\section{Results and discussions} \label{Sec:result}

In this section, we first discuss some generalities pertinent to  the considered processes.
As explained in Appendix A, each exchange process is multiplied by a form factor which contains a cut-off value. In general in the J\"uBo model, those cut-offs are fitted to experimental data and the extracted pole positions depend on the values of those internal parameters.   
In the present study, however, we use the cut-off to postdict
the position of the LHCb pentaquarks without a direct fit to the published line shapes, predicting additional states at the same time.
The exact values of the cut-offs used can be found in Appendix~\ref{App:formalism}. It should be noted that we do not attribute any physical meaning to those model-dependent parameters.
A fit to the experimental data of the $P_c$ states as well as an uncertainty analysis is beyond the scope of this work and postponed to the future.

\begin{figure}[htbp]
    \centering
    \includegraphics[width=0.47\linewidth]{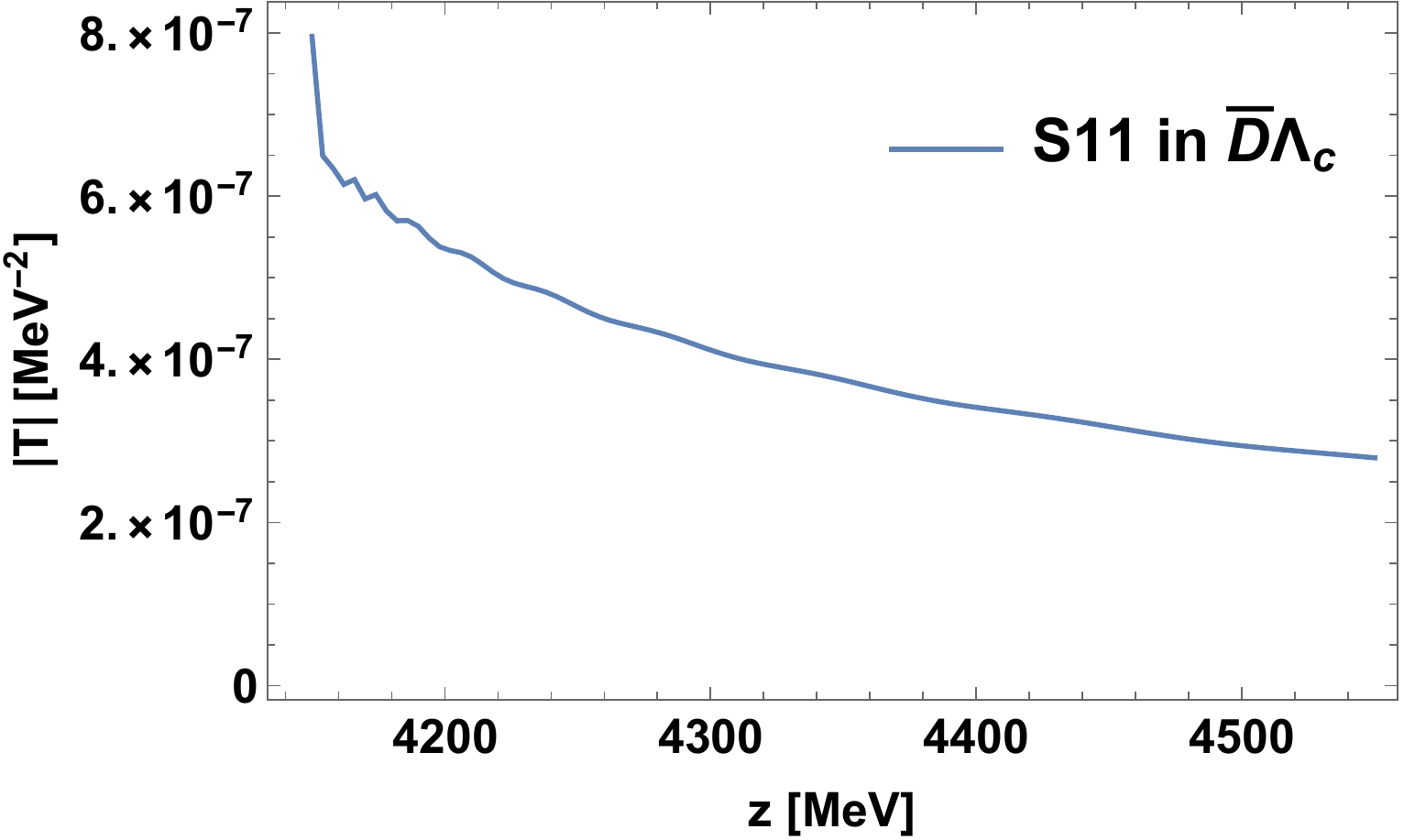} \
    \includegraphics[width=0.47\linewidth]{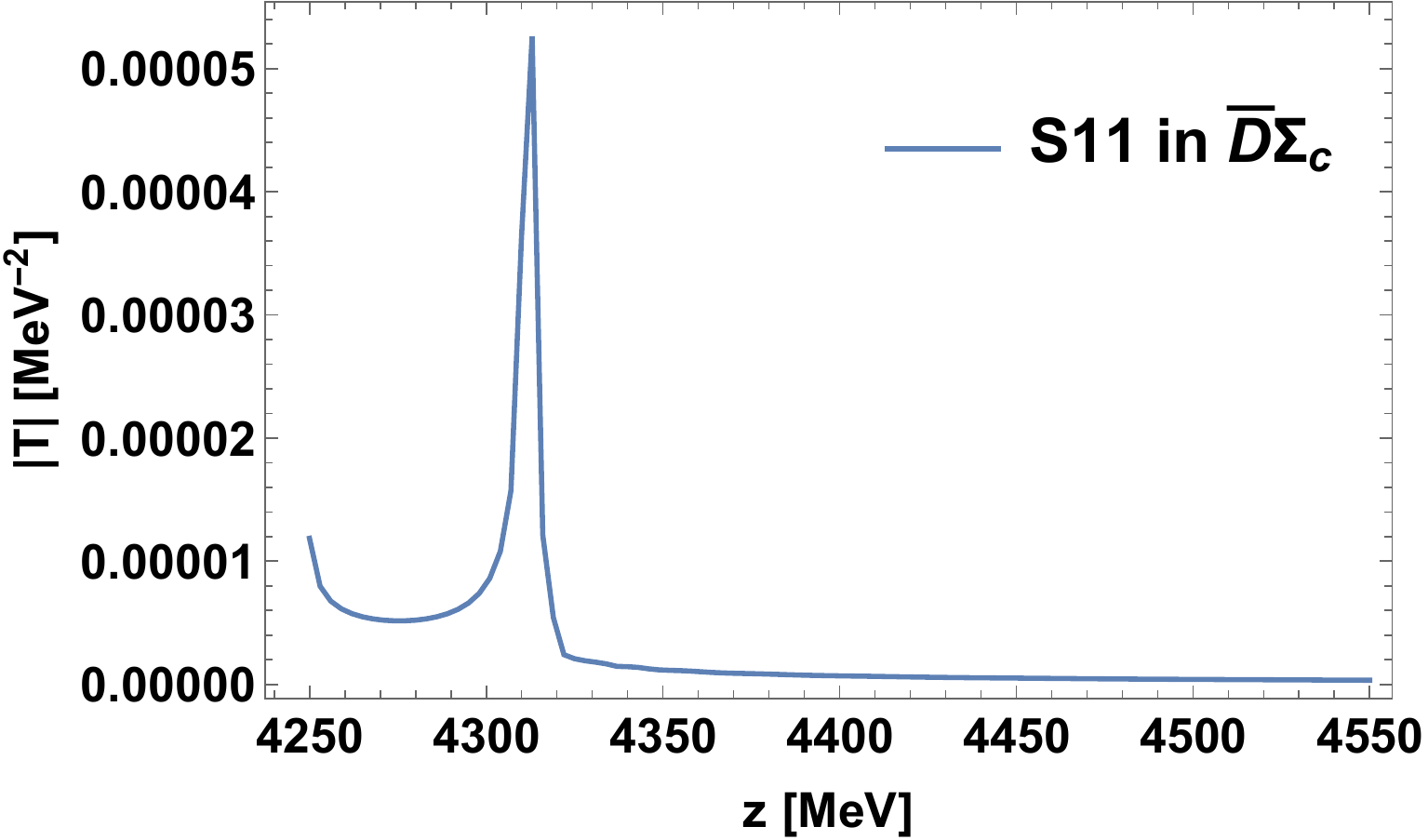} \\
    \includegraphics[width=0.47\linewidth]{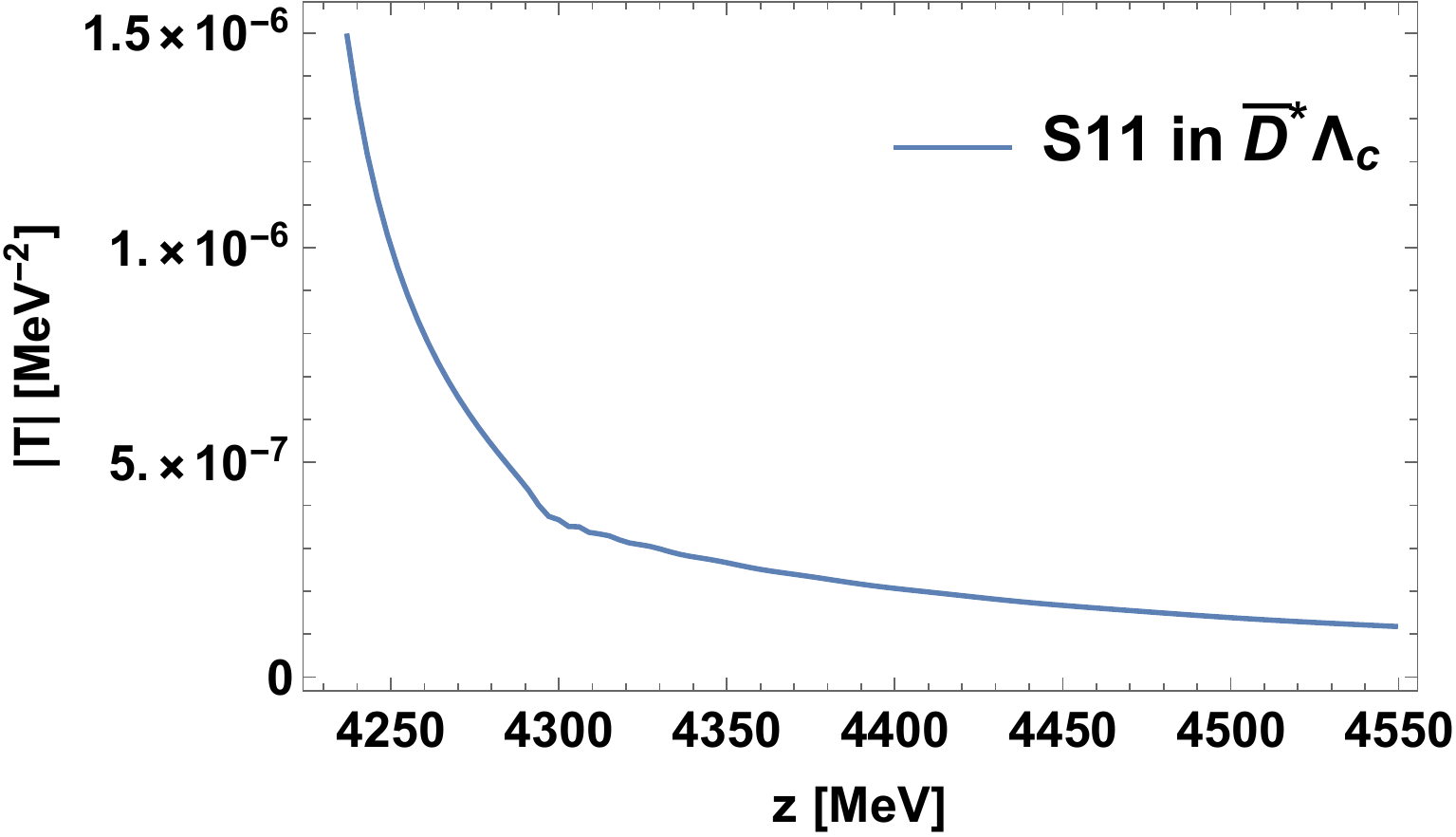} \
    \includegraphics[width=0.47\linewidth]{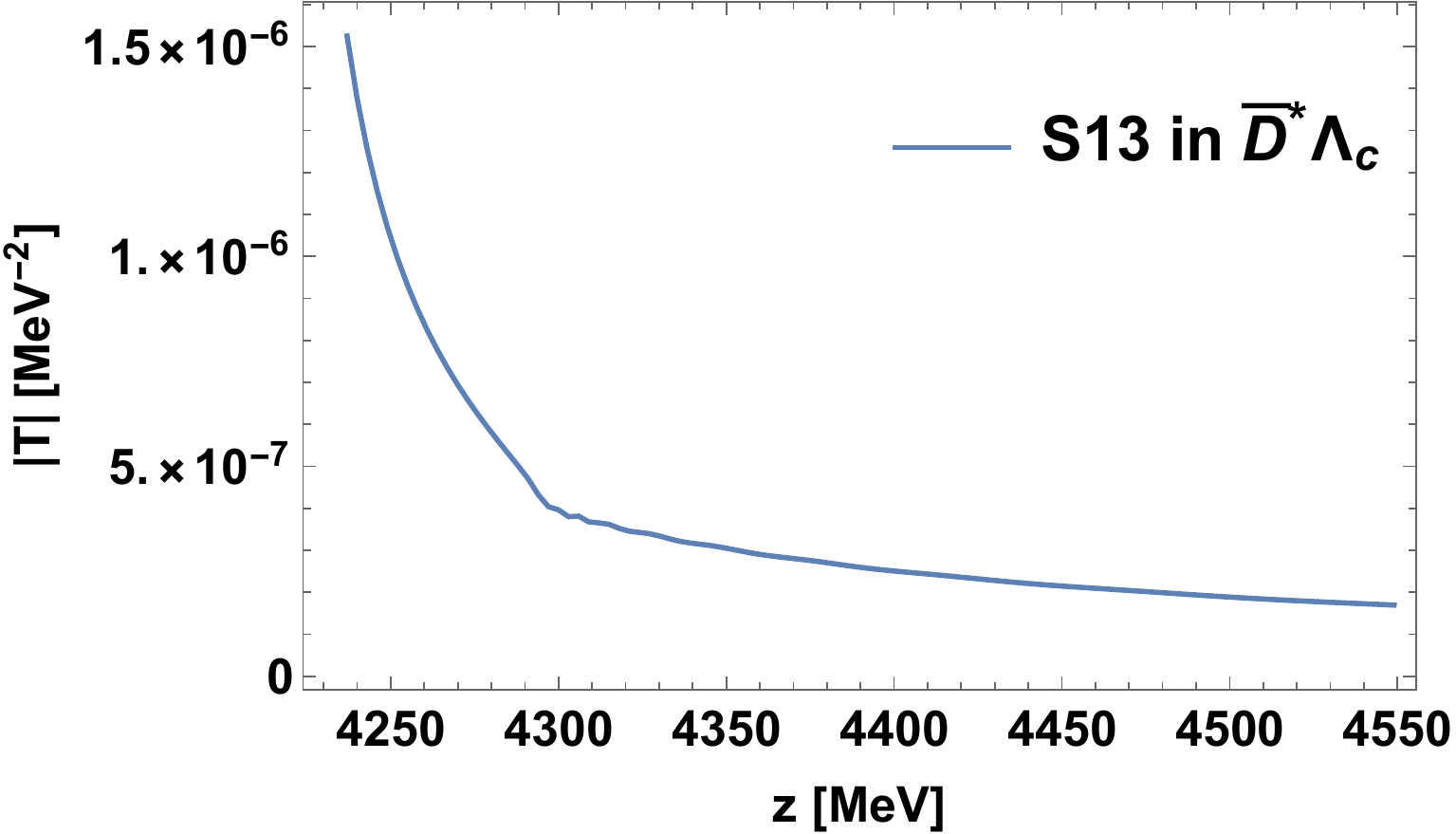} \\
    \includegraphics[width=0.47\linewidth]{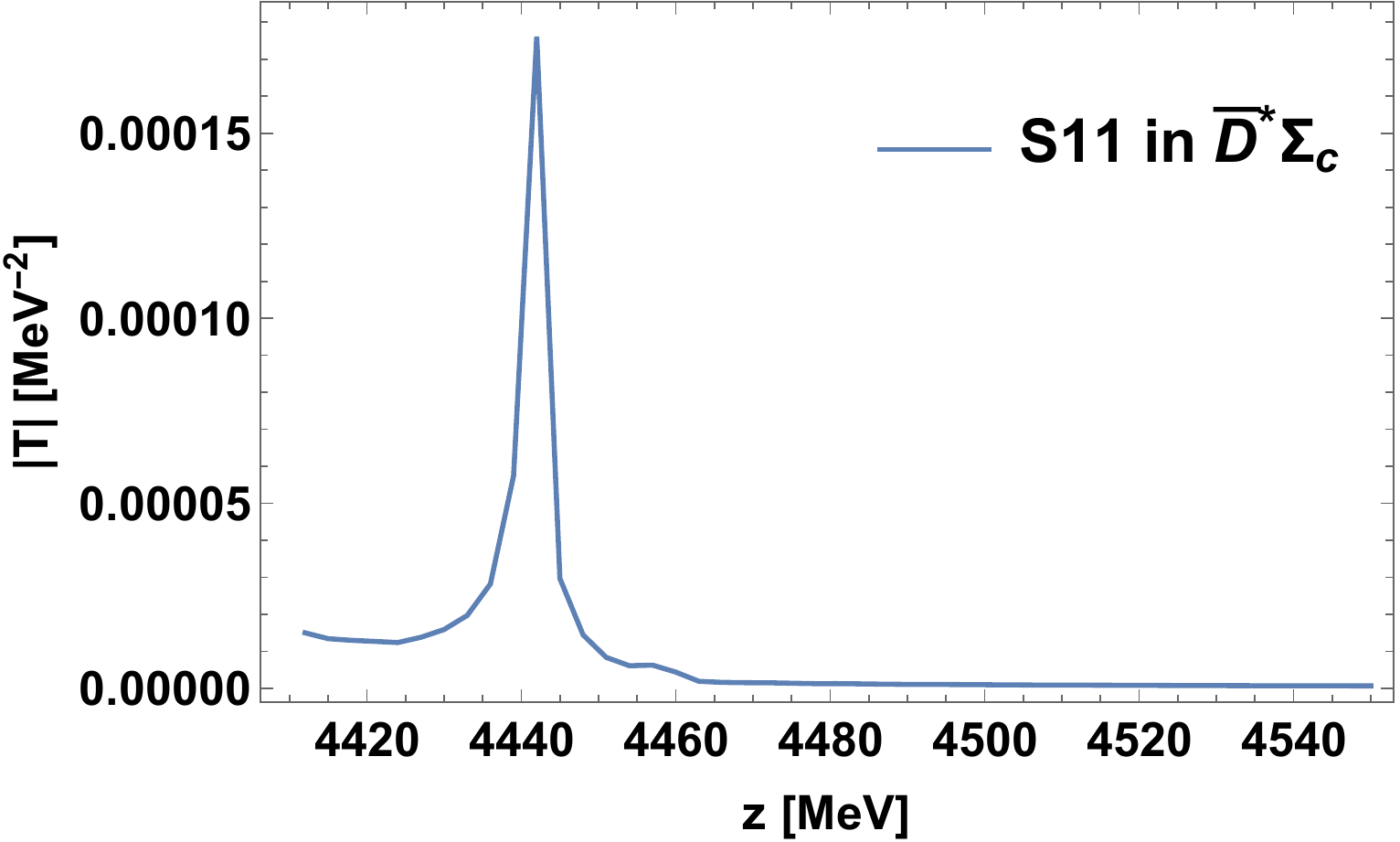} \
    \includegraphics[width=0.47\linewidth]{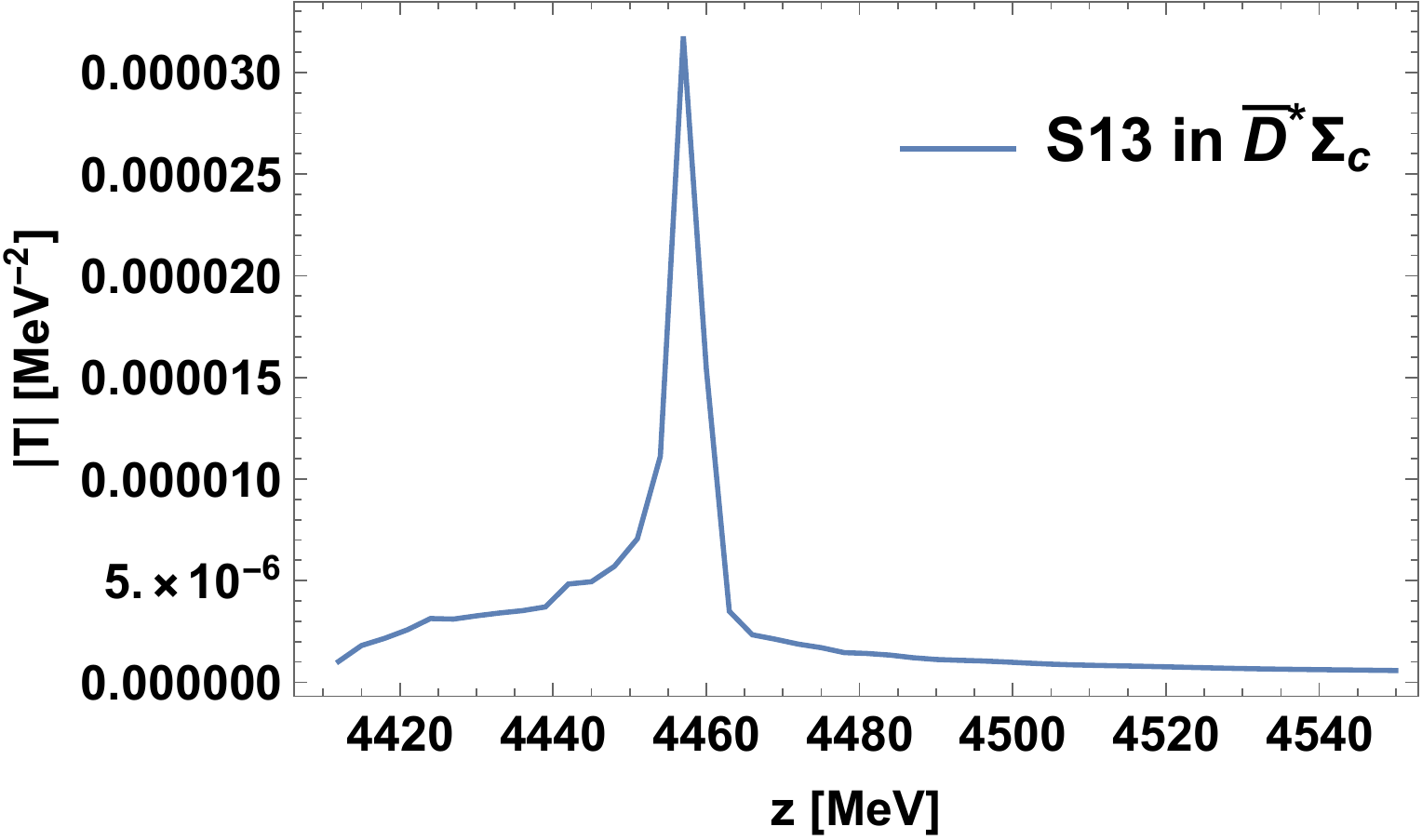} \\
    \includegraphics[width=0.47\linewidth]{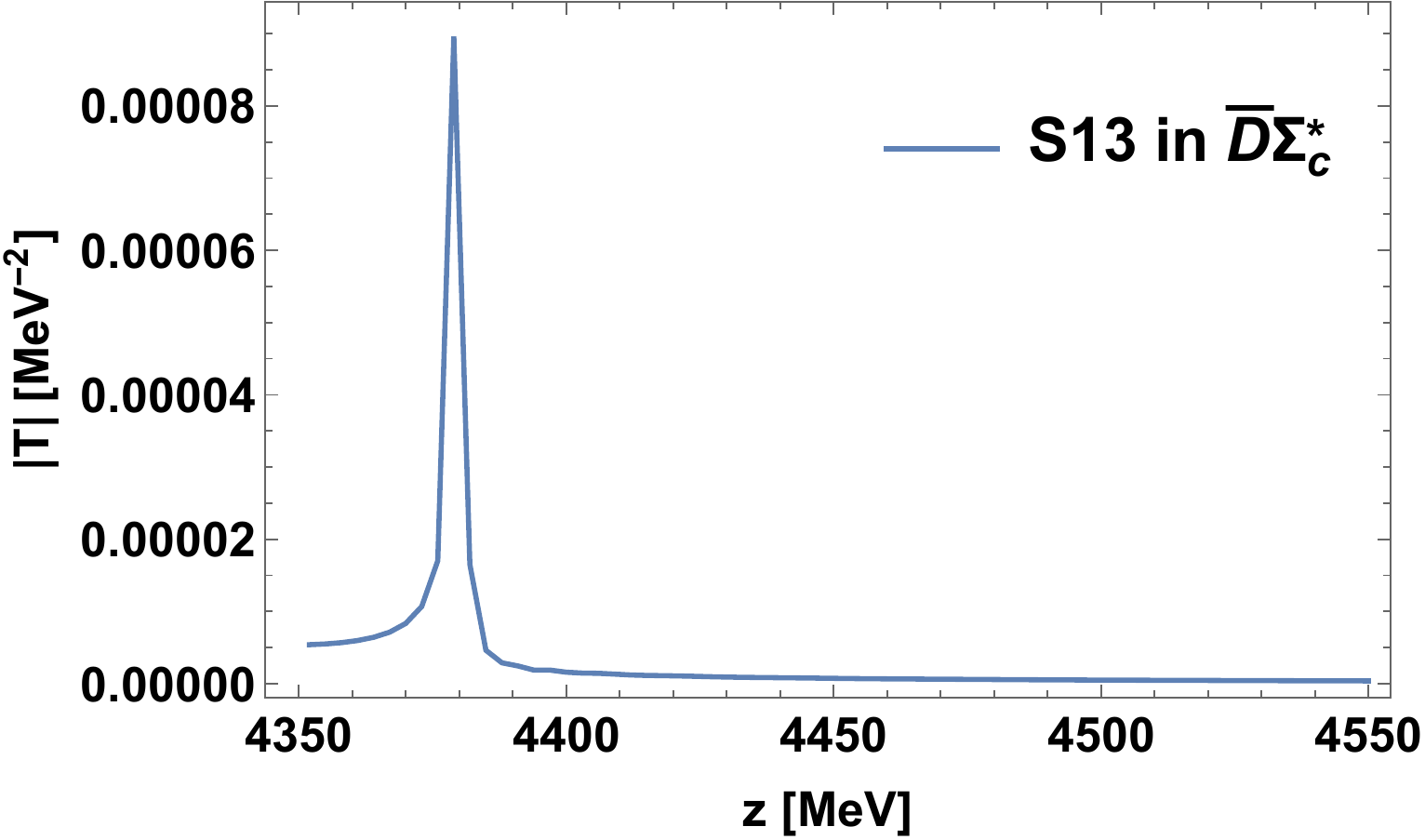}
    \caption{The absolute value of the $T$-matrix for the single channel calculation of different channels in $S$-wave. \label{Fig:t22} }
\end{figure}

We find that  the potentials of $\bar{D}\Sigma_c \to \bar{D}\Sigma_c$, $\bar{D}^*\Sigma_c \to
\bar{D}^*\Sigma_c$ and $\bar{D}\Sigma_c^* \to \bar{D}\Sigma_c^*$ are attractive, and $S$-wave bound states
can be formed in these channels.
To confirm this, we first do the single channel calculation for all the considered channels as a test.
The absolute value of the $T$-matrix for different channels in the single channel case, such as $\bar{D}\Sigma_c \to \bar{D}\Sigma_c$, in $S$-wave is shown in Fig.~\ref{Fig:t22}.
A peak structure around $4310$~MeV is located just below the $\bar{D}\Sigma_c$ threshold.
Similarly, when performing a single channel calculation, we find a peak for
the $\bar{D} \Sigma_c^*$ channel in the $S_{13}$ partial wave.
For the $\bar{D}^*\Sigma_c$ channel, there exist two peaks in the
$S_{11}$ and $S_{13}$ partial waves, respectively.
Contrary to this, the interactions of $\bar{D}\Lambda_c \to \bar{D}\Lambda_c$ and $\bar{D}\Lambda_c^*
\to \bar{D}\Lambda_c^*$ are repulsive indicating that no bound states can be formed.
%

We observe that the effect of the $u$-channel diagrams are negligible  due to the large mass of the
exchanged doubly charmed baryon.
Moreover, for the $t$-channel diagrams, vector meson exchange is more important
than pseudoscalar meson exchange in generating bound states for $\bar{D}^{(*)} \Sigma_c^{(*)} \to
\bar{D}^{(*)} \Sigma_c^{(*)}$.
Except for the two partial waves $S_{11}$ and $S_{13}$ mentioned above, we have not observed any
peak structures  in other partial waves in the single channel case.
%
%

\begin{figure}[htbp]
    \centering
    \includegraphics[width=\linewidth]{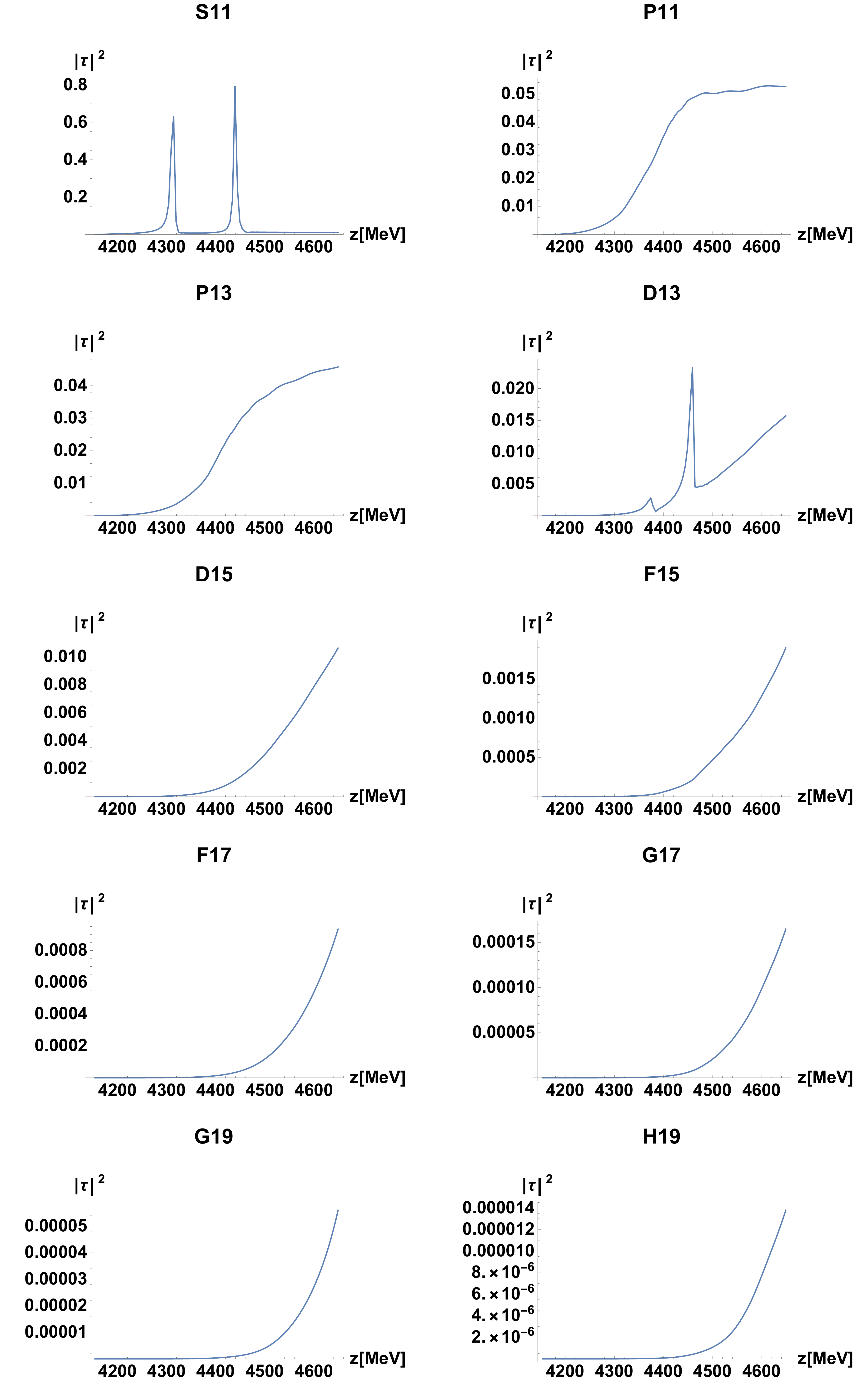}
    \caption{Partial-wave amplitudes squared for $\bar{D}\Lambda_c \to \bar{D}\Lambda_c$ of the full coupled-channel model.
      \label{Fig:tau}}
\end{figure}

We now turn to the results of the complete calculation with all five channels included.
The amplitudes squared of partial waves up to $J=7/2$ for the $\bar{D}\Lambda_c \to \bar{D}\Lambda_c$
process are shown in Fig.~\ref{Fig:tau}.
The relation of the dimensionless partial-wave amplitude $\tau_{\mu\nu}$ to the scattering
amplitude $T_{\mu\nu}$ of Eq.(\ref{Eq:LSEq}) is:
\begin{equation}
    \tau_{\mu \nu} = -\pi \sqrt{\rho_\mu \rho_\nu} T_{\mu \nu},
\end{equation}
where  the phase factor is $\rho_\mu=k_\mu E_\mu \omega_\mu/z$, with $k_\mu$, $E_\mu$ and $\omega_\mu$
the on-shell three-momentum, baryon energy and meson energy of channel $\mu$, respectively.

We need to point out that although some amplitudes are still increasing within the
range we plot, all of them will eventually fall with increasing  energy.

In Fig.~\ref{Fig:tau}, two peaks can be observed in the $S_{11}$ partial wave.
One is located just below the threshold of $\bar{D}\Sigma_c$ and the other one is below the
threshold of $\bar{D}^*\Sigma_c$. Based on their positions, a reasonable assumption is that they are $S$-wave
$\bar{D}\Sigma_c$ and $\bar{D}^*\Sigma_c$ bound states. Likewise, there are two peaks in the $D_{13}$
wave located below the thresholds of $\bar{D}\Sigma_c^*$ and $\bar{D}^*\Sigma_c$, respectively.
Thus, a similar conjecture is that they are the bound states of these two channels.
It should be pointed out that here $D_{13}$ is the name of the partial wave for $\bar{D} \Lambda_c$
channel meaning $J^P=3/2^-$. This name can be different when talking about another channel
for the same state, while $J^P$ is always the same, c.f. Table~\ref{Tab:QuNum}. If not specifically pointed out, we use
the name of the $\bar{D} \Lambda_c$ partial wave.
For example, if the lower observed $J^P=3/2^-$ state in the $D_{13}$ partial wave is a
$\bar{D}\Sigma_c^*$ bound state, it could be an $S$- or $D$-wave bound state.

\begin{figure}[htbp]
    \centering
    \includegraphics[width=0.75\linewidth]{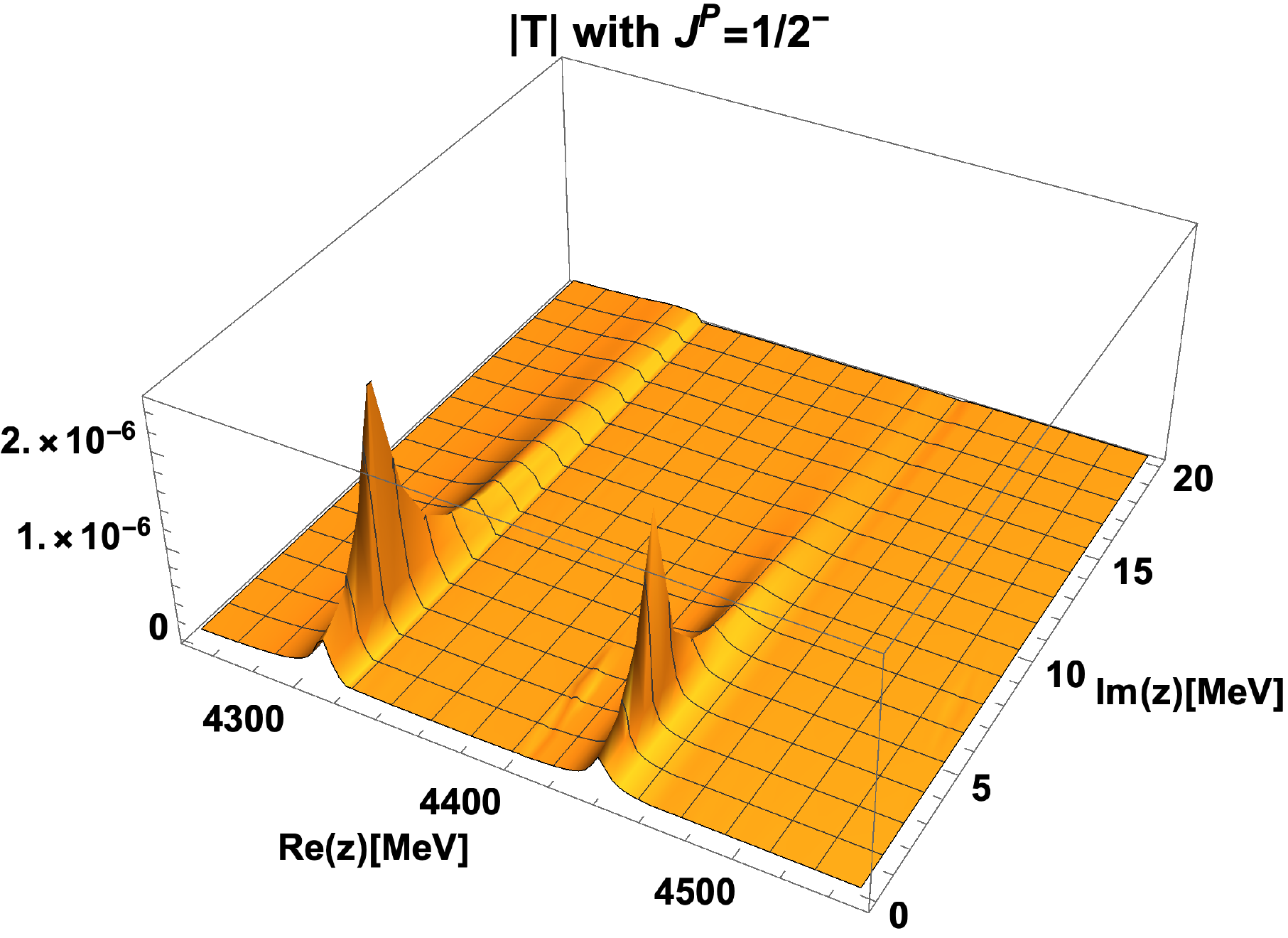} \\
    \includegraphics[width=0.75\linewidth]{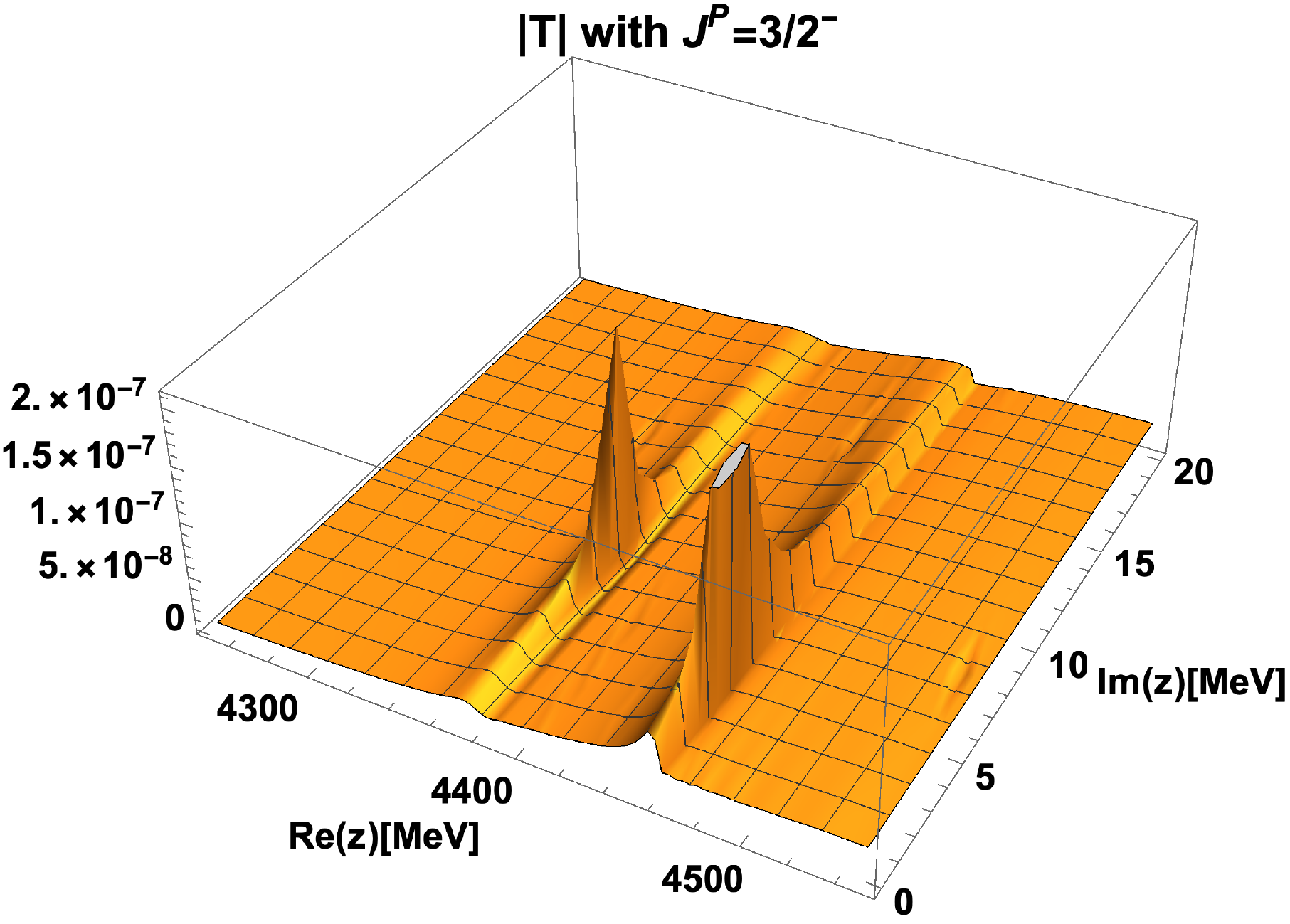}
    \caption{Absolute value $|T|$ of $\bar{D}\Lambda_c \to \bar{D}\Lambda_c$. The upper one is
      the $S_{11}$ wave and the lower one is the $D_{13}$ partial wave. \label{Fig:3d}}
\end{figure}

In order to substantiate the above inferences and clarify the situation in the other partial
waves, we extend the scattering matrix $T$ to the complex energy plane and perform a search
for  dynamically generated poles on the second Riemann sheet. Note that in the present study no ``genuine" $s$-channel pole diagrams are included. For details of the pole searching and the determination of the
coupling strength, i.e. the residue, of a pole with respect to different channels, we refer to Refs.~\cite{Doring:2009yv,Ronchen:2012eg,Shen:2017ayv}.
In Fig.~\ref{Fig:3d}, we present the absolute value $|T|$ of $\bar{D}\Lambda_c \to
\bar{D}\Lambda_c$ for the  $S_{11}$ and $D_{13}$ partial waves  in the complex energy plane.
Two peak structures are clearly visible in both cases, which are in accordance with the
observed peaks in Fig.~\ref{Fig:tau}.

The pole positions and their couplings to different channels  are listed in Table~\ref{Tab:pole}.
It can be seen that all these four poles are very narrow with the widths between 5 to 15~MeV.
The pole at $z_R=4312.43-i2.9$~MeV is about 8~MeV below the threshold
of $\bar{D} \Sigma_c$ and it couples strongly to this channel.
Thus, it can be considered as a $\bar{D} \Sigma_c$ bound state with $J^P=1/2^-$,
and it could be assigned to the $P_c(4312)$ state.
Among all the couplings of the pole at $z_R=4375.89-i7.6$~MeV,
its coupling to the $\bar{D} \Sigma_c^*$ channel is the strongest.
We regard it as a $\bar{D} \Sigma_c^*$ bound state with $J^P=3/2^-$,
where the binding energy is about 10~MeV.
The positions of the remaining two poles are below the threshold
of $\bar{D}^* \Sigma_c$, and their couplings with this channel is the strongest.
These two poles are both viewed as the $S$-wave $\bar{D}^* \Sigma_c$ bound states
with spin-parity $J^P=1/2^-$ and $J^P=3/2^-$, respectively.
We assign them to the observed $P_c(4440)$ and $P_c(4457)$ states.
Note, however, the exact positions of these states do depend on the cut-offs employed in our calculations. 
Furthermore we would like to point out that conjectures about the nature
of the observed states are beyond the scope of the present work.

\begin{table*}[htbp]
    \centering
    \renewcommand\arraystretch{1.3}
    \caption{Pole positions $z_R$, spin-parities $J^P$ and couplings $g_\mu$ for the
      states in $S_{11}$ and $D_{13}$ with $I=\frac12$. \label{Tab:pole}}
        \begin{tabular}{p{2.1cm}<{\centering}|*{4}{p{3.5cm}<{\centering}}}
            \hline
            \hline
             & \multicolumn{4}{c}{$z_R (J^P)$} \\
            \cline{2-5}
            \makecell*{$g_\mu$\\$[10^{-3}{\rm MeV}^{-\frac12}]$} & $4312.43-i2.9 (\frac12^-)$ & $4375.89-i7.6 (\frac32^-)$ & $4439.36-i2.8 (\frac12^-)$  & $4460.00-i3.9 (\frac32^-)$ \\
            \hline
            $g_1$ & $1.2+i0.15$ & $0.23-i0.014$ & $0.96+i0.026$ & $0.25-i0.30$ \\
            $g_2$ & $7.7+i1.0$ & $0.21+i0.25$ & $-0.84+i0.29$ & $-0.15+i0.82$ \\
            $g_3$ & $0.023-i0.43$ & $0.053+i0.82$ & $1.2\cdot 10^{-3}-i0.33$ & $-0.13+i0.16$ \\
            $g_4$ & $0$& $0.065+i0.77$ & $0$ & $-0.13+i0.054$ \\
            $g_5$ & $0.022+i0.24$ & $-0.30+i2.5$ & $0.032+i0.067$ & $0.13-i7.3\cdot 10^{-3}$ \\
            $g_6$ & $-1.3+i6.8$ & $-0.95+i1.3$ & $-0.62+i11.4$ & $0.055+i0.070$ \\
            $g_7$ & $0$ & $1.7+i8.0$ & $0$ & $-0.027-i0.13$ \\
            $g_8$ & $-1.3-i12$ & $-1.9+i3.9$ & $-0.10+i0.48$ & $0.061-i2.1$ \\
            $g_9$ & $0$ & $0.20+i0.023$ & $0$ & $-1.1-i1.5$ \\
            $g_{10}$ & $0.17+i0.49$ & $10.1+i3.6$ & $-0.081-i0.021$ & $0.38-i0.85$ \\
            \hline
            \hline
        \end{tabular}
\end{table*}

For other partial waves, no structures are observed directly
close to the real axis (c.f. Fig.~\ref{Fig:tau}), while we do find several poles in those partial waves
located farther away from the real axis.  All the poles obtained in this study
have isospin $I=1/2$, and no states with isospin $3/2$ are found.
In Table~\ref{Tab:other_poles}, we present the pole positions and spin-parities for some of the observed states, which can be regarded as resonances in $P$- or higher partial waves with relatively large widths.
%
%
The state at $4339.3$~MeV having $J^P=1/2^+$ is close to the mass of the newly observed
$P_c(4337)$ state. Although the width of this state here is larger than
that of the $P_c(4337)$ state, which is about 29~MeV, the situation may change
after adjusting the cut-offs to the experimental data.
Similarly, the state at $4386.0$~MeV with $J^P=5/2^+$, which also
has a large width, could be related to the broad $P_c(4380)$ state proposed in 2015,
where the preferred spin is $3/2$ or $5/2$~\cite{LHCb:2015yax}.
In Ref.~\cite{LHCb:2019kea}, the existence of $P_c(4380)$ is weakened but not ruled out.
Furthermore, we see inconclusive indications for  two poles with even larger
imaginary parts in the $D_{13}$ and $D_{15}$ partial waves, which are not listed
in Table~\ref{Tab:other_poles}.

\begin{table}[htbp]
    \centering
    \renewcommand\arraystretch{1.5}
    \caption{Pole positions $z_R$ and spin-parities $J^P$ for the states in other partial waves.
      Some poles with even larger widths are not listed.\label{Tab:other_poles}}
    \begin{tabular}{p{1.1cm}<{\centering}|p{2.9cm}<{\centering}}
        \hline
        \hline
        $J^P$ & $z_R$[MeV] \\
        \hline
        $\frac12^+$ & $4339.3-i106.3$ \\
        \hline
        $\frac32^+$ & $4401.4-i128.8$ \\
        \hline
        $\frac32^+$ & $4463.1-i90.1$ \\
        \hline
        $\frac52^+$ & $4386.0-i95.2$ \\
        \hline
        $\frac72^-$ & $4430.8-i214.1$ \\
        \hline
        \hline
    \end{tabular}
\end{table}

\section{Summary} \label{Sec:summary}

Based on the previous exploratory study, we extend the J\"ulich-Bonn dynamical coupled-channel
model to the hidden charm sector with more channels taken into account. The channels included
here are $\bar{D} \Lambda_c$, $\bar{D} \Sigma_c$, $\bar{D}^* \Lambda_c$, $\bar{D}^* \Sigma_c$,
$\bar{D} \Sigma_c^*$. Predictions of partial-wave amplitudes are provided with possible
dynamically generated poles examined. A series of bound states and resonances that can
be assigned to the hidden-charm pentaquarks found by LHCb with different spin and parity are generated.
Four $S$-wave bound states with isospin $I=1/2$, whose widths are between 5 to 15~MeV, are
found in the mass range of the experimentally observed $P_c$ states. The $\bar{D} \Sigma_c$
bound state at $z_R=4312.43-i2.9$~MeV with $J^P=1/2^-$ can be assigned to the $P_c(4312)$ state.
The one at $z_R=4375.89-i7.6$~MeV is regarded as a $\bar{D}\Sigma_c^*$ bound state with
$J^P=3/2^-$, since its coupling with this channel is the strongest, and its binding energy is
about 10~MeV. The other two $S$-wave states are treated as the $\bar{D}^*\Sigma_c$ bound states
with one having $J^P=1/2^-$ and the other having $J^P=3/2^-$. They are considered as the
experimentally observed $P_c(4440)$ and $P_c(4457)$ states. In higher partial waves, several
additional resonances with different spin and parity are also dynamically generated.
However, their decay widths are very large and need further confirmation by
fitting the free parameters of the model to experimental data. Such work is underway.
%

\section*{Acknowledgments}

We thank Yu-Fei Wang, Feng-Kun Guo, Jia-Jun Wu, Yu Lu and Fei Huang for useful discussions and valuable comments.
This work is supported by the NSFC and the Deutsche Forschungsgemeinschaft (DFG, German Research
Foundation) through the funds provided to the Sino-German Collaborative
Research Center TRR110 “Symmetries and the Emergence of Structure in QCD”
(NSFC Grant No. 12070131001, DFG Project-ID 196253076 - TRR 110), by the NSFC
Grant No.11835015, No.12047503,  by the Chinese Academy of Sciences (CAS) under Grant No.XDB34030000,
by  the  Chinese  Academy  of Sciences (CAS) President's  International  Fellowship  Initiative
(PIFI)  (grant  no.    2018DM0034) and by  VolkswagenStiftung (grant no.  93562).

\begin{appendix}

\section{Exchange potentials} \label{App:formalism}

\begin{figure}[htbp]
    \subfigure[$\vec{q}=\vec{p}_1-\vec{p}_3$]{
        \includegraphics[width=0.4\linewidth]{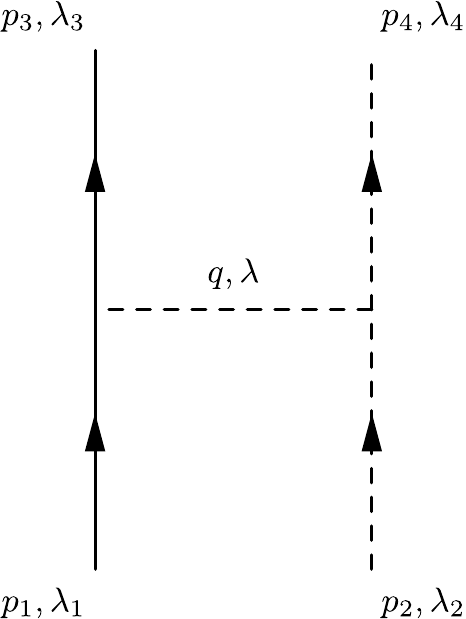}}
    \ \
    \subfigure[$\vec{q}=\vec{p}_1-\vec{p}_4$]{
        \includegraphics[width=0.4\linewidth]{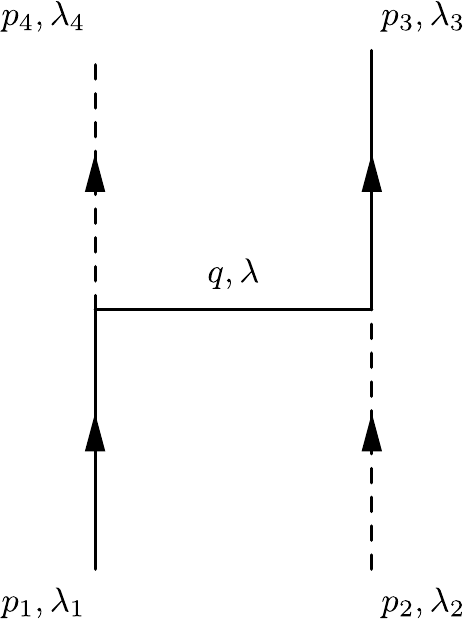}}
    \caption{$t$- and $u$-channel exchange processes. \label{Fig:tuchannel}}
\end{figure}

We consider $t$-channel meson and $u$-channel baryon exchanges in this work.
The corresponding Feynman diagrams are shown in Fig.~\ref{Fig:tuchannel}, where the labeling
of the particles and the momenta are specified.

For each exchange diagram, there is a kinematic normalization factor, an isospin factor (IF) and
form factors $F_1$, $F_2$ in addition to the pseudo-potential ${\cal V}$.
The kinematic normalization factor is $N = \frac1{(2 \pi)^{3}} \frac1{2 \sqrt{\omega_2 \omega_4}}$.
The isospin factors (IF) is calculated using
\begin{align}
  IF(I) =& \sum_{m_1,m_2,m_3,m_4} \langle I_1 m_1 I_2 m_2 | I I_z \rangle \langle I_3 m_3 I_4 m_4 |
  I I_z \rangle \nonumber \\
  &\times \langle I_3 m_3 I_4 m_4 | M^{iso} | I_1 m_1 I_2 m_2 \rangle,
\end{align}
where the notations and details can be found in Ref.~\cite{Gasparyan:PhDthesis}.
All the isospin factors for involved processes are given in Table~\ref{Tab:if}.
Note that the baryon-first convention is always applied in the isospin calculation.
At each vertex, a form factor is applied, its specific form is given by
\begin{equation}\label{eq:ff}
        F(q)=\left(\frac{\Lambda^{2}-m_{e x}^{2}}{\Lambda^{2}+\vec{q}^{2}}\right)^{n},
\end{equation}
where $m_{ex}$, $\vec{q}$ and $\Lambda$ are the mass, the three-momentum of the exchanged particle
and the cut-off, respectively. The powers $n=1, 2$ correspond to monopole or dipole form factors.
For the $t$-channel pseduscalar and vector meson exchange, we use a dipole form factor,
i.e. $n=2$, while a monopole form factor is applied for $u$-channel $\Xi_{cc}$ exchange, i.e. $n=1$.
The transition amplitude is then expressed by
\begin{equation} \label{Eq:potential}
    V_{\lambda_{1}, \lambda_{2}, \lambda_{3}, \lambda_{4}}^{\vec{p}_{1}, \vec{p}_{2}, \vec{p}_{3}, \vec{p}_{4}} =
    N F_{1} F_{2}(\mathrm{IF}) \mathcal{V}.
\end{equation}
The detailed expressions of all the needed potentials in this work are presented below.

\begin{table}[]
    \centering
    \renewcommand\arraystretch{1.3}
    \caption{The isospin factors for all the involved processes in this work. \label{Tab:if}}
    \begin{tabular}{p{3.3cm}<{\centering}|p{1.8cm}<{\centering}|*{2}{p{1.2cm}<{\centering}}}
        \hline
        \hline
        Process & \makecell*{Exchanged\\Particle} & IF$\left(\frac12\right)$ & IF$\left(\frac32\right)$ \\
        \hline
        $\Lambda_c \bar{D}^{(*)} \to \Lambda_c \bar{D}^{(*)}$ & $\omega$ & 1 & 0 \\
            & $\Xi_{c c}$ & 1 & 0 \\
        $\Lambda_c \bar{D}^{(*)} \to \Sigma_c^{(*)} \bar{D}^{(*)}$ & $\pi/\rho$ & $-\sqrt{3}$ & 0 \\
            & $\Xi_{cc}$ & $\sqrt{3}$ & 0 \\
        $\Sigma_c^{(*)} \bar{D}^{(*)} \to \Sigma_c^{(*)} \bar{D}^{(*)}$ & $\eta / \eta^{\prime} / \omega$ & 1 & 1 \\
            & $\pi / \rho$ & -2 & 1 \\
            & $\Xi_{cc}$ & -1 & 2 \\
        \hline
        \hline
    \end{tabular}
\end{table}

\subsection{Amplitudes} \label{App:potential}

There are in total 15 different kinds of exchange pseudo-potentials ${\cal V}$ considered here.
As clarified in Eq.~(\ref{Eq:potential}), other factors including the normalization factor, isospin factors and form factors have been isolated.
Here, we present the explicit expressions of all the needed pseudo-potentials using the notations
in Fig.~\ref{Fig:tuchannel}.
\begin{widetext}
\begin{description}
    \item[Type 1]: $BP \to BP$ via vector meson exchange,
    \begin{equation}
        {\cal V} = - g_a g_b {\bar u}_3
        \left( \frac{\gamma^\mu - i \frac{\kappa}{2m} \sigma^{\mu \nu} q_{\nu}}{z-\omega_t-\omega_2-E_3}
		+ \frac{\gamma^\mu - i \frac{\kappa}{2m} \sigma^{\mu \nu} \tilde{q}_{\nu}}{z-\omega_t-E_1-\omega_4} \right)
		\frac{(p_2+p_4)_\mu}{2\omega_t} u_1.
    \end{equation}
    \item[Type 2]: $BP \to BP$ via baryon exchange,
    \begin{equation}
    	{\cal V} = \frac{g_a g_b}{m_P^2} {\bar u}_3 \gamma^5 \Slash{p}_2
    	\left( \frac{\Slash{q}+m}{z-E_u-\omega_2-\omega_4} +\frac{\tilde{\Slash{q}}+m}{z-E_u-E_1-E_3} \right)
    	\frac1{2E_u} \gamma^5 \Slash{p}_4 u_1.
    \end{equation}
    \item[Type 3]: $BP\to BV$ via pseudoscalar meson exchange,
    \begin{equation}
		{\cal V} = i \frac{g_a g_b}{m_P} {\bar u}_3 \gamma^5 \left( \frac{\Slash{q} (p_2-q)_\mu}{z-\omega_t-\omega_2-E_3}
		+\frac{\tilde{\Slash{q}} (p_2-\tilde{q})_\mu}{z-\omega_t-E_1-\omega_4} -2\omega_t\gamma^0\delta^0_\mu \right)
		\frac1{2\omega_t} u_1 \epsilon_4^{* \mu}.
    \end{equation}
    \item[Type 4]: $BP\to BV$ via vector meson exchange,
    \begin{equation}
		{\cal V} = \frac{g_a g_b}{m_V} {\bar u}_3
        \left( \frac{\gamma^\mu - i \frac{\kappa}{2m} \sigma^{\mu \nu} q_{\nu}}{z-\omega_t-\omega_2-E_3}
		+ \frac{\gamma^\mu - i \frac{\kappa}{2m} \sigma^{\mu \nu} \tilde{q}_{\nu}}{z-\omega_t-E_1-\omega_4} \right)
		\frac1{2\omega_t} u_1 \epsilon_{\alpha \beta \lambda \mu} p_4^\alpha \epsilon_4^{* \beta} p_2^\lambda.
    \end{equation}
    \item[Type 5]: $BP\to BV$ via baryon exchange,
    \begin{equation}
		{\cal V} = -i \frac{g_a g_b}{m_P} {\bar u}_3 \gamma^5 \Slash{p}_2
		\left( \frac{\Slash{q}+m}{z-E_u-\omega_2-\omega_4} +\frac{\tilde{\Slash{q}}+m}{z-E_u-E_1-E_3} \right)
        \frac1{2E_u} (\gamma^\mu -i\frac{\kappa}{2m} \sigma^{\mu \nu} p_{4 \nu}) \epsilon_{4 \mu}^* u_1.
    \end{equation}
    \item[Type 6]: $BP\to DP$ via vector meson exchange,
    \begin{equation}
		{\cal V} = \frac{g_a g_b}{m_V} {\bar u}_3^{\mu} \gamma^5
		\left( \frac{\gamma_\nu q_\mu - \Slash{q} g_{\mu \nu}}{z-\omega_t-\omega_2-E_3}
		+ \frac{\gamma_\nu \tilde{q}_\mu - \tilde{\Slash{q}} g_{\mu \nu}}{z-\omega_t-E_1-\omega_4} \right)
		\frac{(p_2+p_4)^\nu}{2\omega_t} u_1.
    \end{equation}
    \item[Type 7]: $BP\to DP$ via baryon exchange,
    \begin{equation}
		{\cal V} = \frac{g_a g_b}{m_P^2} {\bar u}_3^{\mu} \left( \frac{\Slash{q}+m}{z-E_u-\omega_2-\omega_4}
		+\frac{\tilde{\Slash{q}}+m}{z-E_u-E_1-E_3} \right) \frac{p_{2 \mu}}{2E_u} \gamma^5 \Slash{p}_4 u_1.
    \end{equation}
    \item[Type 8]: $BV\to BV$ via vector meson exchange,
    \begin{equation}
    	{\cal V} = - g_a g_b {\bar u}_3 \Big[ \frac{(\gamma^\mu - i \frac{\kappa}{2m} \sigma^{\mu \nu} q_{\nu})
    	[\epsilon_2^\nu \epsilon_{4 \mu}^* (p_4+q)_\nu - \epsilon_2^\nu \epsilon_{4 \nu}^* (p_2+p_4)_\mu
    	+ \epsilon_{2 \mu} \epsilon_4^{* \nu} (p_2-q)_\nu] }{z-\omega_t-\omega_2-E_3} + \frac{q \to \tilde{q}}{z-\omega_t-E_1-\omega_4} \Big] \frac1{2\omega_t} u_1.
    \end{equation}
    \item[Type 9]: $BV\to BV$ via pseudoscalar meson exchange,
    \begin{equation}
		{\cal V} = -i \frac{g_a g_b}{m_P m_V} {\bar u}_3 \gamma^5
		\left( \frac{\Slash{q}}{z-\omega_t-\omega_2-E_3} + \frac{\tilde{\Slash{q}}}{z-\omega_t-E_1-\omega_4} \right)
		\frac1{2\omega_t} u_1 \epsilon_{\mu \nu \lambda \tau} p_2^\mu \epsilon_2^\nu p_4^\lambda \epsilon_4^{* \tau}.
    \end{equation}
    \item[Type 10]: $BV\to BV$ via baryon exchange,
	\begin{equation}
		{\cal V} = g_a g_b {\bar u}_3 (\Slash{\epsilon}_2 +i\frac{\kappa}{2m} \sigma^{\mu \nu} p_{2 \nu} \epsilon_{2 \mu})
		\left( \frac{\Slash{q}+m}{z-E_u-\omega_2-\omega_4} +\frac{\tilde{\Slash{q}}+m}{z-E_u-E_1-E_3} \right) \frac1{2E_u} (\Slash{\epsilon}_4^* -i\frac{\kappa}{2m} \sigma^{\alpha \beta} p_{4 \beta} \epsilon_{4 \alpha}^*) u_1.
	\end{equation}
    \item[Type 11]: $BV\to DP$ via pseudoscalar meson exchange,
    \begin{equation}
		{\cal V} = i \frac{g_a g_b}{m_P} {\bar u}_3^{\mu} \left( \frac{q_\mu (p_4+q)_\nu}{z-\omega_t-\omega_2-E_3}
		+ \frac{\tilde{q}_\mu (p_4+\tilde{q})_\nu}{z-\omega_t-E_1-\omega_4}
		+ 2\omega_t \delta_\mu^0 \delta_\nu^0 \right) \frac1{2\omega_t} u_1 \epsilon_2^\nu.
    \end{equation}
    \item[Type 12]: $BV\to DP$ via vector meson exchange,
    \begin{equation}
		{\cal V} = \frac{g_a g_b}{m_V^2} {\bar u}_{3,\mu} \gamma^5
		\left( \frac{(\gamma^\beta q^\mu - \Slash{q} g^{\mu \beta})q^\alpha}{z-\omega_t-\omega_2-E_3}
		+ \frac{(\gamma^\beta \tilde{q}^\mu - \tilde{\Slash{q}} g^{\mu \beta})\tilde{q}^\alpha}{z-\omega_t-E_1-\omega_4} \right)
		\frac1{2\omega_t} u_1 \epsilon_{\alpha \beta \lambda \tau} p_2^\lambda \epsilon_2^\tau.
    \end{equation}
    \item[Type 13]: $BV\to DP$ via baryon exchange,
    \begin{equation}
		{\cal V} = i \frac{g_a g_b}{m_P m_V} {\bar u}_3^{\mu} \gamma^5 (p_{2 \mu}\Slash{\epsilon}_2 - \epsilon_{2 \mu} \Slash{p}_2)
		\left( \frac{\Slash{q}+m}{z-E_u-\omega_2-\omega_4} +\frac{\tilde{\Slash{q}}+m}{z-E_u-E_1-E_3} \right) \frac1{2E_u} \gamma^5 \Slash{p}_4 u_1.
    \end{equation}
    \item[Type 14]: $DP\to DP$ via vector meson exchange,
    \begin{equation}
		{\cal V} = g_a g_b {\bar u}_3^{\tau}
		\left( \frac{\gamma^\mu - i \frac{\kappa}{2m} \sigma^{\mu \nu} q_{\nu}}{z-\omega_t-\omega_2-E_3}
		+ \frac{\gamma^\mu - i \frac{\kappa}{2m} \sigma^{\mu \nu} \tilde{q}_{\nu}}{z-\omega_t-E_1-\omega_4} \right)
		\frac{(p_2+p_4)_\mu}{2\omega_t} u_{1,\tau}.
    \end{equation}
    \item[Type 15]: $DP\to DP$ via baryon exchange,
    \begin{equation}
		{\cal V} = \frac{g_a g_b}{m_P^2} {\bar u}_3^{\mu} \left( \frac{\Slash{q}+m}{z-E_u-\omega_2-\omega_4}
		+ \frac{\tilde{\Slash{q}}+m}{z-E_u-E_1-E_3} \right)	\frac{p_{2 \mu} p_{4 \nu}}{2E_u} u_1^{\nu}.
    \end{equation}
\end{description}
\end{widetext}
Here, $q$ is the exchanged momentum in the first time ordering and $\tilde{q}$ means the
second time ordering with $\tilde{q}^0=-q^0$.
The indices 1 and 3 (2 and 4) denote the incoming and outgoing baryon (meson).
$E_i$ and $\omega_i$ are the on-shell energies for a baryon and a meson, respectively.
A ``type'' number is introduced to identify which potential each process corresponds to.

In Table~\ref{Tab:sum}, we list all the processes with the exchanged particle considered
in the calculation, and the corresponding ``type'' of each specific process together with
the coupling constants $g_{a,b}$ and cut-off values in the form factor are also given.
Note that, for simplicity, we take one cut-off for each process, not for each vertex.
For the $t$-channel pseudoscalar meson exchange and $u$-channel baryon exchange, we take
a uniform value for the same exchanged particle. We adjust the cut-offs for the $t$-channel
vector meson exchanges to obtain positions of the dynamically generated poles near
the experimentally observed $P_c$ states. As mentioned above, here we emphasize again that our
numerical results depend on the cut-off values, and the results given in this work are for
this specific set of cut-offs.
In the cases where vector mesons couple to octet baryons and vector mesons couple to decuplet
baryons, there is also a tensor coupling $f_b=g_b \kappa$, which is not listed in Table~\ref{Tab:sum},
but should be consistent with the vector coupling $g_b$.
The explicit relations of the coupling constants $g$ are given below.

\begin{table}[htbp]
    \centering
    \renewcommand\arraystretch{1.16}
    \caption{The types of potentials, the couplings and the cut-offs in the form factor
      for all the considered processes with the exchanged particle (Ex). \label{Tab:sum}}
    \begin{tabular}{c|c|c|cc|c}
        \hline
        \hline
        Process & Ex & Type & $g_a$ & $g_b$ & $\Lambda$ [MeV] \\
        \hline
        $\Lambda_c \bar{D} \to \Lambda_c \bar{D}$ & $\omega$ & 1 & $g_{D D \omega}$ & $g_{\Lambda_c \Lambda_c \omega}$ & 2800 \\
         & $\Xi_{cc}$ & 2 & $g_{\Xi_{cc} \Lambda_c D}$ & $g_{\Xi_{cc} \Lambda_c D}$ & 3000 \\
        $\Lambda_c \bar{D} \to \Sigma_c \bar{D}$ & $\rho$ & 1 & $g_{D D \rho}$ & $g_{\Sigma_c \Lambda_c \rho}$ & 2800 \\
         & $\Xi_{cc}$ & 2 & $g_{\Xi_{cc} \Sigma_c D}$ & $g_{\Xi_{cc} \Lambda_c D}$ & 3000 \\
        $\Lambda_c \bar{D} \to \Lambda_c \bar{D}^*$ & $\omega$ & 4 & $g_{D^* \omega D}$ & $g_{\Lambda_c \Lambda_c \omega}$ & 1500 \\
         & $\Xi_{cc}$ & 5 & $g_{\Xi_{cc} \Lambda_c D}$ & $g_{\Xi_{cc} \Lambda_c D^*}$ & 3000 \\
        $\Lambda_c \bar{D} \to \Sigma_c \bar{D}^*$ & $\pi$ & 3 & $g_{D \pi D^*}$ & $g_{\Sigma_c \Lambda_c \pi}$ & 800 \\
         & $\rho$ & 4 & $g_{D^* \rho D}$ & $g_{\Sigma_c \Lambda_c \rho}$ & 1800 \\
         & $\Xi_{cc}$ & 5 & $g_{\Xi_{cc} \Sigma_c D}$ & $g_{\Xi_{cc} \Lambda_c D^*}$ & 3000 \\
        $\Lambda_c \bar{D} \to \Sigma_c^* \bar{D}$ & $\rho$ & 6 & $g_{D D \rho}$ & $g_{\Sigma_c^* \Lambda_c \rho}$ & 1800 \\
         & $\Xi_{cc}$ & 7 & $g_{\Sigma_c^* \Xi_{cc} D}$ & $g_{\Xi_{cc} \Lambda_c D}$ & 3000 \\
        $\Sigma_c \bar{D} \to \Sigma_c \bar{D}$ & $\rho$ & 1 & $g_{D D \rho}$ & $g_{\Sigma_c \Sigma_c \rho}$ & 2700 \\
         & $\omega$ & 1 & $g_{D D \omega}$ & $g_{\Sigma_c \Sigma_c \omega}$ & 2700 \\
         & $\Xi_{cc}$ & 2 & $g_{\Xi_{cc} \Sigma_c D}$ & $g_{\Xi_{cc} \Sigma_c D}$ & 3000 \\
        $\Sigma_c \bar{D} \to \Lambda_c \bar{D}^*$ & $\pi$ & 3 & $g_{D \pi D^*}$ & $g_{\Sigma_c \Lambda_c \pi}$ & 800 \\
         & $\rho$ & 4 & $g_{D^* \rho D}$ & $g_{\Sigma_c \Lambda_c \rho}$ & 1500 \\
         & $\Xi_{cc}$ & 5 & $g_{\Xi_{cc} \Lambda_c D}$ & $g_{\Xi_{cc} \Sigma_c D^*}$ & 3000 \\
        $\Sigma_c \bar{D} \to \Sigma_c \bar{D}^*$ & $\pi$ & 3 & $g_{D \pi D^*}$ & $g_{\Sigma_c \Sigma_c \pi}$ & 800 \\
         & $\eta$ & 3 & $g_{D \eta D^*}$ & $g_{\Sigma_c \Sigma_c \eta}$ & 1000 \\
         & $\eta^\prime$ & 3 & $g_{D \eta^\prime D^*}$ & $g_{\Sigma_c \Sigma_c \eta^\prime}$ & 1000 \\
         & $\rho$ & 4 & $g_{D^* \rho D}$ & $g_{\Sigma_c \Sigma_c \rho}$ & 1500 \\
         & $\omega$ & 4 & $g_{D^* \omega D}$ & $g_{\Sigma_c \Sigma_c \omega}$ & 1500 \\
         & $\Xi_{cc}$ & 5 & $g_{\Xi_{cc} \Sigma_c D}$ & $g_{\Xi_{cc} \Sigma_c D^*}$ & 3000 \\
        $\Sigma_c \bar{D} \to \Sigma_c^* \bar{D}$ & $\rho$ & 6 & $g_{D D \rho}$ & $g_{\Sigma_c^* \Sigma_c \rho}$ & 1500 \\
         & $\omega$ & 6 & $g_{D D \omega}$ & $g_{\Sigma_c^* \Sigma_c \omega}$ & 1500 \\
         & $\Xi_{cc}$ & 7 & $g_{\Sigma_c^* \Xi_{cc} D}$ & $g_{\Xi_{cc} \Sigma_c D}$ & 3000 \\
        $\Lambda_c \bar{D}^* \to \Lambda_c \bar{D}^*$ & $\omega$ & 8 & $g_{D^* D^* \omega}$ & $g_{\Lambda_c \Lambda_c \omega}$ & 2000 \\
         & $\Xi_{cc}$ & 10 & $g_{\Xi_{cc} \Lambda_c D^*}$ & $g_{\Xi_{cc} \Lambda_c D^*}$ & 3000 \\
        $\Lambda_c \bar{D}^* \to \Sigma_c \bar{D}^*$ & $\pi$ & 9 & $g_{D^* D^* \pi}$ & $g_{\Sigma_c \Lambda_c \pi}$ & 800 \\
         & $\rho$ & 8 & $g_{D^* D^* \rho}$ & $g_{\Sigma_c \Lambda_c \rho}$ & 3000 \\
         & $\Xi_{cc}$ & 10 & $g_{\Xi_{cc} \Sigma_c D^*}$ & $g_{\Xi_{cc} \Lambda_c D^*}$ & 3000 \\
        $\Lambda_c \bar{D}^* \to \Sigma_c^* \bar{D}$ & $\pi$ & 11 & $g_{D \pi D^*}$ & $g_{\Sigma_c^* \Lambda_c \pi}$ & 800 \\
         & $\rho$ & 12 & $g_{D^* \rho D}$ & $g_{\Sigma_c^* \Lambda_c \rho}$ & 1300 \\
         & $\Xi_{cc}$ & 13 & $g_{\Sigma_c^* \Xi_{cc} D^*}$ & $g_{\Xi_{cc} \Lambda_c D}$ & 3000 \\
        $\Sigma_c \bar{D}^* \to \Sigma_c \bar{D}^*$ & $\pi$ & 9 & $g_{D^* D^* \pi}$ & $g_{\Sigma_c \Sigma_c \pi}$ & 800 \\
         & $\eta$ & 9 & $g_{D^* D^* \eta}$ & $g_{\Sigma_c \Sigma_c \eta}$ & 1000 \\
         & $\eta^\prime$ & 9 & $g_{D^* D^* \eta^\prime}$ & $g_{\Sigma_c \Sigma_c \eta^\prime}$ & 1000 \\
         & $\rho$ & 8 & $g_{D^* D^* \rho}$ & $g_{\Sigma_c \Sigma_c \rho}$ & 2600 \\
         & $\omega$ & 8 & $g_{D^* D^* \omega}$ & $g_{\Sigma_c \Sigma_c \omega}$ & 1000 \\
         & $\Xi_{cc}$ & 10 & $g_{\Xi_{cc} \Sigma_c D^*}$ & $g_{\Xi_{cc} \Sigma_c D^*}$ & 3000 \\
        $\Sigma_c \bar{D}^* \to \Sigma_c^* \bar{D}$ & $\pi$ & 11 & $g_{D \pi D^*}$ & $g_{\Sigma_c^* \Sigma_c \pi}$ & 800 \\
         & $\eta$ & 11 & $g_{D \eta D^*}$ & $g_{\Sigma_c^* \Sigma_c \eta}$ & 1000 \\
         & $\eta^\prime$ & 11 & $g_{D \eta^\prime D^*}$ & $g_{\Sigma_c^* \Sigma_c \eta^\prime}$ & 1000 \\
         & $\rho$ & 12 & $g_{D^* \rho D}$ & $g_{\Sigma_c^* \Sigma_c \rho}$ & 1500 \\
         & $\omega$ & 12 & $g_{D^* \omega D}$ & $g_{\Sigma_c^* \Sigma_c \omega}$ & 1500 \\
         & $\Xi_{cc}$ & 13 & $g_{\Sigma_c^* \Xi_{cc} D^*}$ & $g_{\Xi_{cc} \Sigma_c D}$ & 3000 \\
        $\Sigma_c^* \bar{D} \to \Sigma_c^* \bar{D}$ & $\rho$ & 14 & $g_{D D \rho}$ & $g_{\Sigma_c^* \Sigma_c^* \rho}$ & 1700 \\
         & $\omega$ & 14 & $g_{D D \omega}$ & $g_{\Sigma_c^* \Sigma_c^* \omega}$ & 1700 \\
         & $\Xi_{cc}$ & 15 & $g_{\Sigma_c^* \Xi_{cc} D}$ & $g_{\Sigma_c^* \Xi_{cc} D}$ & 3000 \\
        \hline
        \hline
    \end{tabular}
\end{table}

\subsection{Coupling constants}\label{App:coupling}

The coupling constants for all the vertices in the calculation are related to
each other by SU(3) and SU(4) flavor symmetry~\cite{deSwart:1963pdg,Okubo:1975sc}.
The employed relations of the various coupling constants are given by
the following expressions:
\begin{itemize}
	\item Couplings for pseudoscalar meson, pseudoscalar meson and vector meson:
	\begin{align}\label{Eq:gPPV}
		g_{D D \rho}    &=   g_{PPV},		\nonumber \\
		g_{D D \omega}    &=   g_{PPV},		\nonumber \\
		g_{D \pi D^*}    &= - g_{PPV},		\nonumber \\
		g_{D \eta D^*}   &=  - \frac1{\sqrt{3}} g_{PPV},		\nonumber \\
		g_{D \eta^\prime D^*}  &= - \sqrt{\frac23} g_{PPV},		\nonumber \\
		g_{D D J/\psi}   &= - \sqrt{2} g_{PPV},
	\end{align}
	where $g_{PPV}=3.018$.
	\item Couplings for vector meson, vector meson and pseudoscalar meson:
	\begin{align}\label{Eq:gVVP}
		g_{D^* \rho D}   &=   \frac1{\sqrt{2}} g_{VVP},		\nonumber \\
		g_{D^* \omega D}   &=   \frac1{\sqrt{2}} g_{VVP},		\nonumber \\
		g_{D^* D^* \pi}   &=   \frac1{\sqrt{2}} g_{VVP},		\nonumber \\
		g_{D^* D^* \eta}  &=   \frac1{\sqrt{6}} g_{VVP},		\nonumber \\
		g_{D^* D^* \eta^\prime} &=   \frac1{\sqrt{3}} g_{VVP},		\nonumber	\\
		g_{D^* J/\psi D}  &=   g_{VVP},
	\end{align}
	where $g_{VVP}=-7.070$.
	\item Couplings for vector meson, vector meson and vector meson:
	\begin{align}\label{Eq:gVVV}
		g_{D^* D^* \rho}  &= - \frac1{\sqrt{2}} g_{VVV},		\nonumber \\
		g_{D^* D^* \omega}  &= - \frac1{\sqrt{2}} g_{VVV},		\nonumber \\
		g_{D^* D^* J/\psi} &=   g_{VVV},
	\end{align}
	where $g_{VVV}=2.298$.
    \item Couplings for octet baryon, octet baryon and pseudoscalar meson:
    \begin{align}\label{Eq:gBBP}
    	g_{\Lambda_c N D}  &=  - \frac{3\sqrt{3}}5 g_{BBP},		\nonumber \\
    	g_{\Sigma_c N D}   &=  \frac15 g_{BBP},\nonumber \\
    	g_{\Sigma_c \Lambda_c \pi} &=  \frac{2\sqrt{3}}5 g_{BBP},		\nonumber \\
    	g_{\Sigma_c \Sigma_c \pi}  &=  \frac45 g_{BBP},		\nonumber \\
    	g_{\Sigma_c \Sigma_c \eta} &=  \frac4{5\sqrt{3}} g_{BBP},		\nonumber \\
    	g_{\Sigma_c \Sigma_c \eta^\prime}  &=  \frac{4\sqrt{2}}{5\sqrt{3}} g_{BBP},		\nonumber \\
    	g_{\Xi_{cc} \Lambda_c D}   &=  \frac{\sqrt3}5 g_{BBP},		\nonumber \\
        g_{\Xi_{cc} \Sigma_c D}   &=  - g_{BBP},
    \end{align}
    where $g_{BBP}=0.989$.
	\item Vector couplings for octet baryon, octet baryon and vector meson:
	\begin{align}\label{Eq:gBBV}
		g_{\Lambda_c N D^*}     &= - \frac1{\sqrt{3}} g_{BBV} (1 + 2 \alpha_{BBV}),		\nonumber \\
	    g_{\Sigma_c N D^*}     &=   g_{BBV} (1 - 2 \alpha_{BBV}),		\nonumber \\
	    g_{\Lambda_c \Lambda_c \omega}   &=   \frac23 g_{BBV} (5 \alpha_{BBV} - 2),		\nonumber \\
	    g_{\Lambda_c \Lambda_c J/\psi}  &=   \sqrt{2} g_{BBV},		\nonumber \\
	    g_{\Sigma_c \Lambda_c \rho}   &=   \frac2{\sqrt{3}} g_{BBV} (1 - \alpha_{BBV}),		\nonumber \\
	    g_{\Sigma_c \Sigma_c \rho}   &=   2 g_{BBV} \alpha_{BBV},		\nonumber \\
	    g_{\Sigma_c \Sigma_c \omega}   &=   2 g_{BBV} \alpha_{BBV},		\nonumber \\
	    g_{\Sigma_c \Sigma_c J/\psi}  &=   \sqrt{2} g_{BBV},		\nonumber	\\
	    g_{\Xi_{cc} \Lambda_c D^*}  &=   \frac1{\sqrt{3}} g_{BBV} (4 \alpha_{BBV} - 1),		\nonumber \\
	    g_{\Xi_{cc} \Sigma_c D^*}  &= - g_{BBV},
	\end{align}
	where $g_{BBV}=3.249$, $\alpha_{BBV}=1.15$.
    \item Tensor couplings for octet baryon, octet baryon and vector meson:
    \begin{align}\label{Eq:fBBV}
    	f_{\Lambda_c N D^*}     &= - \frac1{2\sqrt{3}} f_{N N \omega} - \frac{\sqrt{3}}{2} f_{N N \rho},		\nonumber	\\
    	f_{\Sigma_c N D^*}     &= - \frac12 f_{N N \omega} + \frac12 f_{N N \rho},		\nonumber \\
    	f_{\Lambda_c \Lambda_c \omega}   &=   \frac56 f_{N N \omega} - \frac12 f_{N N \rho},		\nonumber \\
    	f_{\Lambda_c \Lambda_c J/\psi}  &=  0,		\nonumber	\\
    	f_{\Sigma_c \Lambda_c \rho}   &= - \frac1{2\sqrt{3}} f_{N N \omega} + \frac{\sqrt{3}}{2} f_{N N \rho},		\nonumber \\
    	f_{\Sigma_c \Sigma_c \rho}   &=   \frac12 f_{N N \omega} + \frac12 f_{N N \rho},		\nonumber \\
    	f_{\Sigma_c \Sigma_c \omega}   &=   \frac12 f_{N N \omega} + \frac12 f_{N N \rho},		\nonumber \\
    	f_{\Sigma_c \Sigma_c J/\psi}  &=   0,	\nonumber	\\
    	f_{\Xi_{cc} \Lambda_c D^*}  &=   0,	\nonumber	\\
    	f_{\Xi_{cc} \Sigma_c D^*}  &=   0,
    \end{align}
    where $f_{NN\rho}=19.819$, $f_{NN\omega}=0$.
	\item Couplings for decuplet baryon, octet baryon and pseudoscalar meson:
	\begin{align}\label{Eq:gBDP}
		g_{\Sigma_c^* N D}      &= - \frac1{\sqrt{6}} g_{BDP},		\nonumber \\
		g_{\Sigma_c^* \Lambda_c \pi}    &=   \frac1{\sqrt{2}} g_{BDP},		\nonumber \\
		g_{\Sigma_c^* \Sigma_c \pi}    &=   \frac1{\sqrt{6}} g_{BDP},		\nonumber \\
		g_{\Sigma_c^* \Sigma_c \eta}   &=   \frac1{2\sqrt{2}} g_{BDP},		\nonumber \\
		g_{\Sigma_c^* \Sigma_c \eta^\prime}  &=   \frac13 g_{BDP},		\nonumber \\
		g_{\Sigma_c^* \Xi_{cc} D}   &= - \frac1{\sqrt{6}} g_{BDP},
	\end{align}
	where $g_{BDP}=2.127$.
    \item Couplings for decuplet baryon, octet baryon and vector meson:
    \begin{align}\label{Eq:gBDV}
    	g_{\Sigma_c^* N D^*}     &= - \frac1{\sqrt{6}} g_{BDV},		\nonumber \\
    	g_{\Sigma_c^* \Lambda_c \rho}   &=   \frac1{\sqrt{2}} g_{BDV},		\nonumber \\
    	g_{\Sigma_c^* \Sigma_c \rho}   &=  \frac1{\sqrt{6}} g_{BDV},		\nonumber \\
    	g_{\Sigma_c^* \Sigma_c \omega}   &= - \frac1{\sqrt{6}} g_{BDV},		\nonumber \\
    	g_{\Sigma_c^* \Sigma_c J/\psi}  &= - \frac1{\sqrt{3}} g_{BDV},		\nonumber \\
    	g_{\Sigma_c^* \Xi_{cc} D^*}  &= - \frac1{\sqrt{6}} g_{BDV},
    \end{align}
    where $g_{BDV}=16.031$.
	\item Vector and tensor couplings for decuplet baryon, decuplet baryon and vector meson:
	\begin{align}\label{Eq:gDDV}
		g_{\Sigma_c^* \Sigma_c^* \rho}  &=   2 g_{DDV},		\nonumber \\
		g_{\Sigma_c^* \Sigma_c^* \omega}  &=   2 g_{DDV},		\nonumber \\
		f_{\Sigma_c^* \Sigma_c^* \rho}  &=   \frac12 \kappa_\rho g_{DDV},		\nonumber \\
		f_{\Sigma_c^* \Sigma_c^* \omega}  &=   \frac12 \kappa_\rho g_{DDV},
	\end{align}
	where $g_{DDV}=7.674$ and $\kappa_{\rho}=6.1$~\cite{Mergell:1995bf}.
\end{itemize}

\end{appendix}

\bibliographystyle{plain}

\end{document}